# Elementary Monte Carlo model of the anisotropic recrystallization and "anti-ripening" under intensive stirring and high supersaturations.

Serhii Abakumov[(1)], Eugene Rabkin[(2)], Andriy Gusak[(1,3,c)]

[(1)] Department of Physics, Cherkasy National University, 18000, Ukraine
[(2)] Faculty of Materials Science and Engineering, Technion – Israel Institute of Technology, 3200003 Haifa, Israel
[(3)] Ensemble3 Centre of Excellence, Warsaw, 01-919, Poland
[c)] corresponding author, amgusak@ukr.net

## I. Introduction

Vanadium pentaoxide $V_2O_5$ is a material with unique electronic structure finding numerous applications in energy conversion and storage [1]. Recently, a new method of fibrous (rod-like or/and belt-like) oxide ($TiO_2$, $V_2O_5$) nanostructures fabrication was suggested involving intensive stirring of commercial $V_2O_5$ powder in saline water at elevated or even at room temperatures [2-6]. This process can be interpreted in the framework of the theory of open driven systems with appropriate ballistic terms [7–10]. Yet, recent experiments demonstrated that fibrous $V_2O_5$ structures form without stirring as well (though in this case the process takes years instead of days).

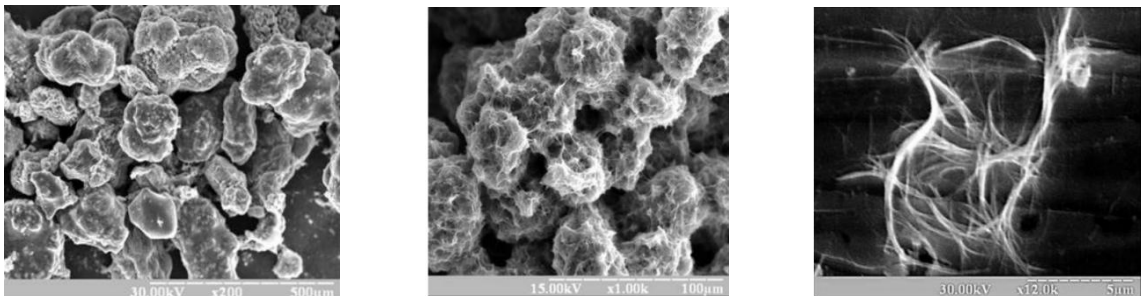

*Fig. 1. General picture of $V_2O_5$ fibers and belts formation under stirring [6].*

In the cited works [1-6], the recrystallization in open anisotropic system with high supersaturations due to intensive stirring was observed. The shape of obtained crystal is very far from the equilibrium crystal shape given by Wulff construction, proving that the conditions of recrystallization are far from equilibrium. Moreover, these structures do not demonstrate any tendency to relaxation towards equilibrium shapes. In particular, these structures do not show any tendency to minimization of surface energies ("antiripening"). It may mean that the obtained structures are metastable. Alternatively, it may mean that the open systems under the action of stationary external factors may tend to highly anisotropic (for example, fibrous) steady-state shapes.

We are trying to construct some simple model which can explain the formation of fibrous morphology from commercial more-or-less equiaxial powder not only due to intensive stirring but as well in still water (though very slowly in this case).

Our first attempts [4-6] were based on the very simplified and linearized Onsager-type phenomenological model, treating all facets as flat and taking the velocity of each facet movement as proportional to corresponding thermodynamic force. According to the concept of ballistic effects in the driven system [6-10], we introduced the additional, ballistic velocity of detachments (shrinking) determined by the stirring intensity and different for different facets. Ballistic detachments increase the supersaturation and lead to additional erosion of non-favorable facets, with simultaneous recrystallization of the favorable facets. This should lead to additional elongation along the direction $(010)$. This approach had two main drawbacks: (1) it did not take into account the non-linear effects at high super- or under-saturations; (2) it did not take into account the terrace-ledge-kink (TLK) structure of the real facets.

We try to improve (at least, partially) the mentioned drawbacks. For this we use the Monte Carlo model with standard Metropolis algorithm, taking into account three basic elementary processes - thermally activated attachment, thermally activated detachment, ballistic detachment, with probability independent on temperature but dependent on the stirring intensity. At that, in this MC approach, we try to neglect the ionic nature of attaching and detaching clusters, and approximate the process as consisting of attaching and detaching the neutral elementary anisotropic "bricks" which we will call "quasiatoms".

In Section II we formulate the basic model.

In Section III we try to simulate and understand the nucleation and growth of more-or less isotropic or highly anisotropic structures at various driving forces. We find the interval of driving forces (or equivalent supersaturations) within which the forming structures may nucleate homogeneously and become very anisotropic. Heterogeneous nucleation (starting from preexisting structure) also may lead to formation of elongated particle.

In Section IV we study the behavior of nanoparticle at constant negative driving force (undersaturation). We show that shrinkage may lead to formation of very rough surfaces and even to fibrous structure at micro-level.

In Section V we study the essentially nonlinear dependence between driving force (here constant in time) and average velocity of crystal growth (at positive driving force) or shrinkage (at negative driving force).

In Section VI we introduce the crystal growth within the limited volume of surrounding medium, so that now the feedback from medium is taken into account. For example, attachment of one atom to the crystal means the depletion of surrounding medium, with respective decrease of driving force. Vice versa, each detachment increases the concentration and the corresponding driving force.

In Sections VII and VIII we introduce the ballistic detachments with probability depending on stirring intensity. In Section VII this probability is taken as isotropic, and in Section VIII the ballistic probability is made dependent on the surface energy change due to detachment. Both Sections give similar results for self-adjustment of the preexisting cluster in the limited medium. We show the possibility of elongated particles formation due to ballistic detachment from $(100)$ and from $(001)$ facets with consequent attachment to $(010)$ face.

In Section IX we introduce the ensemble of preexisting clusters in limited volume with account of ripening (including competition via the mean-field not only between the existing particles, but as well between the different facets of these particles).

## II. Model

$V_2O_5$ has an orthorhombic structure (Space group: Pmmn) with primitive cell containing 14 atoms, in the form of rectangular parallelepiped containing 14 atoms, with sizes
$$a_x = 5.85 \text{ Å}, a_y = 1.78 \text{ Å}, a_z = 4.41 \text{ Å},$$
and with anisotropic surface tensions calculated in [3] with account of surface reconstruction and Van der Waals interactions:
$$\gamma(100) = \gamma(yz) = 0.41 \text{ J/m}^2$$
$$\gamma(010) = \gamma(zx) = 0.54 \text{ J/m}^2$$
$$\gamma(001) = \gamma(xy) = 0.22 \text{ J/m}^2$$

To apply, say, Metropolis algorithm to the attachment or detachment of elementary cell containing 14 atoms, would be a bad choice since energy change related to cell adding or removal, is too big, in comparison with kT, so that the process becomes deterministic. On the other hand, we are looking for some simplified description, without going into details of individual ions or clusters attaching or detaching the growing crystal.

Therefore, for simplicity, in our model we try to represent $V_2O_5$ as consisting of elementary unit cells containing single averaged anisotropic "quasiatom" with renormalized sizes

$$a_{1x} = \frac{a_x}{14^{\frac{1}{3}}} = 2.42724 \times 10^{-10} \text{ m},$$

$$a_{1y} = \frac{a_y}{14^{\frac{1}{3}}} = 7.38546 \times 10^{-11} \text{ m},$$

$$a_{1z} = \frac{a_z}{14^{\frac{1}{3}}} = 1.82977 \times 10^{-10} \text{ m},$$

and quasiatomic volume
$$\Omega = a_{1x} \cdot a_{1y} \cdot a_{1z} = 3.28010 \times 10^{-30} \text{ m}^3$$

Medium surrounding the crystal, is treated as a lattice gas with some concentration (atomic fraction) Cgas=C of isolated quasiatoms. Chemical potential of quasiatoms in the surrounding medium is
$$\mu(gas) = kT \times \ln C,$$
Considering the $V_2O_5$ crystal as a pure substance built of quasiatoms results in the following expression for the chemical potential of quasiatoms in the crystal (Ccrystal=1):
$$\mu(crystal) = E_{bulk} + kT \cdot \ln 1 = E_{bulk} = -E_{isol}$$
Equilibrium means equality of chemical potentials
$\mu(Cgas = Ceq) = \mu(crystal)$, so that
kT*lnCeq=-Eisol => Ceq=exp(-Eisol/kT)=0.959*10^(-7)
at the temperature of 300 K. Driving force of crystallization per one quasiatom is
$$\Delta\mu = \mu(crystal) - \mu(gas) = -E_{isol} - kT \ln C = -kT \ln\left(\frac{C}{C_{eq}}\right)$$

We can distinguish some characteristic values for bulk driving force and corresponding supersaturations S and concentrations of quasiatoms in surrounding the medium. We classify these characteristic values according to the TLK model (terrace-ledge-kink):

1. Energy necessary to attach (detach) the quasiatom to (from) the planar terraces (001), (100), (010):
$$E_3 = E(terrace\, 001) = 2 \cdot \left(\gamma(010) \cdot a_{1z} \cdot a_{1x} + \gamma(100) \cdot a_{1y} \cdot a_{1z}\right) =$$
$$= \frac{2}{14^{2/3}} \cdot (0.54 \cdot 4.41 \cdot 5.85 + 0.41 \cdot 1.78 \cdot 4.41) = 5.90471 \times 10^{-20}$$

Supersaturation $\quad S_3 = exp\left(\frac{E_3}{kT}\right) = exp\left(\frac{59.05}{1.38 \cdot 3}\right) = 1.55329 \times 10^6$

Corresponding concentration in the medium (fraction of quasiatoms at the sites of lattice gas:

$$C_3 = C_{eq} \cdot exp\left(\frac{E_3}{kT}\right) = 0.953 \times 10^{-7} \cdot 1.56 \times 10^6 = 1.48925 \times 10^{-1}$$

$$E_2 = E(terrace\ 100) = 2 \cdot \left(\gamma(001) \cdot a_{1x} \cdot a_{1y} + \gamma(010) \cdot a_{1z} \cdot a_{1x}\right) =$$
$$= \frac{2}{14^{2/3}} \cdot (0.22 \cdot 5.85 \cdot 1.78 + 0.54 \cdot 4.41 \cdot 5.85) = 5.58535 \times 10^{-20}\ J$$

$$S_2 = exp\left(\frac{E_2}{kT}\right) = exp\left(\frac{55.85}{1.38 \cdot 3}\right) = 7.18443 \times 10^5$$

$$C_2 = C_{eq} \cdot exp\left(\frac{E_2}{kT}\right) = 0.953 \times 10^{-7} \cdot 7.18 \times 10^5 = 6.88819 \times 10^{-2}$$

$$E_1 = E(terrace\ 010) = 2 \cdot \left(\gamma(001) \cdot a_{1x} \cdot a_{1y} + \gamma(100) \cdot a_{1y} \cdot a_{1z}\right) =$$
$$= \frac{2}{14^{2/3}} \cdot (0.22 \cdot 5.85 \cdot 1.78 + 0.41 \cdot 1.78 \cdot 4.41) = 1.89688 \times 10^{-20}\ J$$

$$S_1 = exp\left(\frac{E_1}{kT}\right) = exp\left(\frac{18.97}{1.38 \cdot 3}\right) = 9.74828 \times 10^1$$

$$C_1 = C_{eq} \cdot exp\left(\frac{E_1}{kT}\right) = 0.953 \times 10^{-7} \cdot 9.75 \times 10^1 = 9.34633 \times 10^{-6}$$

2. Energy necessary to attach (detach) the quasiatom to (from) the ledges along X or along Y on the terrace (001), ledges along Y or along Z on the terrace (100), ledges along Z or along X on the terrace (010):

$$E_X = E\big(ledge\ along\ X\ on\ terrace\ (001)\big) = E\big(ledge\ along\ X\ on\ terrace\ (010)\big)$$
$$E_X = 2 \cdot \gamma(100) \cdot a_{1y} \cdot a_{1z} = \frac{2}{14^{2/3}} \cdot (0.41 \cdot 1.78 \cdot 4.41) = 1.10812 \times 10^{-20}\ J$$

$$E_Y = E\big(ledge\ along\ Y\ on\ terrace\ (100)\big) = E\big(ledge\ along\ Y\ on\ terrace\ (001)\big)$$
$$E_Y = 2 \cdot \gamma(010) \cdot a_{1z} \cdot a_{1x} = \frac{2}{14^{2/3}} \cdot (0.54 \cdot 4.41 \cdot 5.85) = 4.79659 \times 10^{-20}\ J$$

$$E_Z = E\big(ledge\ along\ Z\ on\ terrace\ (001)\big) = E\big(ledge\ along\ X\ on\ terrace\ (001)\big)$$
$$E_Z = 2 \cdot \gamma(001) \cdot a_{1x} \cdot a_{1y} = \frac{2}{14^{2/3}} \cdot (0.22 \cdot 5.85 \cdot 1.78) = 7.88757 \times 10^{-21}\ J$$

At that:
$$E_3 = E_Y + E_X; \quad E_2 = E_Z + E_Y; \quad E_1 = E_Z + E_X.$$

Monte Carlo algorithm is based on the cycle, each step of which contains :
(1) one attempt of thermally activated detachment of quasiatom randomly chosen among the surface quasiatoms of the existing cluster – according to Metropolis algorithm,
(2) one attempt of thermally activated attachment of quasiatom in the site neighboring to existing cluster – also according to Metropolis algorithm.
(3) one attempt of ballistic detachment of quasiatom (for details look into Sections VII-IX)

### III. Nucleation and growth at fixed bulk driving force under high supersaturations

In this section we consider the crystallization starting from single quasiatom (homogeneous nucleation) or, alternatively, starting from pre-existing cluster of various sizes (heterogeneous nucleation). At that, our pre-existing cluster may have atomically flat facets or (alternatively) facets with terraces-ledges-kinks structure. We simulate the nucleation and growth under fixed bulk driving force (fixed supersaturation S)

$$\Delta \mu = - kT \ln\left(\frac{C}{C_{eq}}\right) = const < 0$$

We performed simulation in the broad range of driving forces from maximum Eisol (sufficient for stability of single isolated quasiatom) to zero (corresponding to quasiatom fully buried in the bulk). At that, we distinguished, first of all, the characteristic values of energy, corresponding to quasiatom attachment to the flat terraces (001), (100) and (010). It gave us 4 subregions of the energy scale:

$$E_3 < dg_{bulk} = - \Delta\mu < E_{isol}$$
$$E_2 < dg_{bulk} = - \Delta\mu < E_3$$
$$E_1 < dg_{bulk} = - \Delta\mu < E_2$$
$$0 < dg_{bulk} = - \Delta\mu < E_1$$

Inside each of these subregions one can distinguish the characteristic energies corresponding to the formation of ledges.

At each step we make two attempts by Metropolis algorithm –

(1) Attempt of adding one more "brick" (quasiatom) to the already existing crystal cluster at the randomly chosen external face of the existing cluster.

(2) Attempt of removing one "brick" (quasiatom) randomly chosen among the marginal bricks (e.g. having at least one free face).

(3) We stop our attempts when the new formed cluster comes to the walls at least in one dimension (along one axis) or, alternatively, if the cluster dissolves completely, or if the number of attachment/detachment attempts exceeds 1000000.

### III.1. Homogeneous nucleation (starting from one quasiatom)

We start with the first pre-existing quasiatom in the center (homogeneous nucleation). Note that the growing cluster is shown in the 3D space with different scale of the axes $X, Y, Z$. Namely, lengths in these directions correspond not to absolute size but instead to the number of periods $X/a_{1x}$, $Y/a_{1y}$, $Z/a_{1z}$. Thus, ideal cube in our picture corresponds to rectangular parallelepiped elongated along X and compressed along Y (about 3 times) in real space.

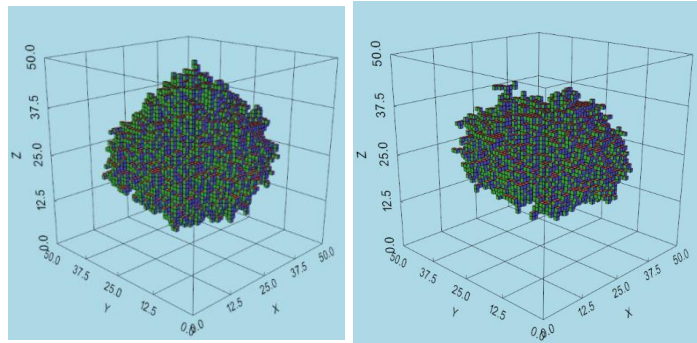

(a)          (b)

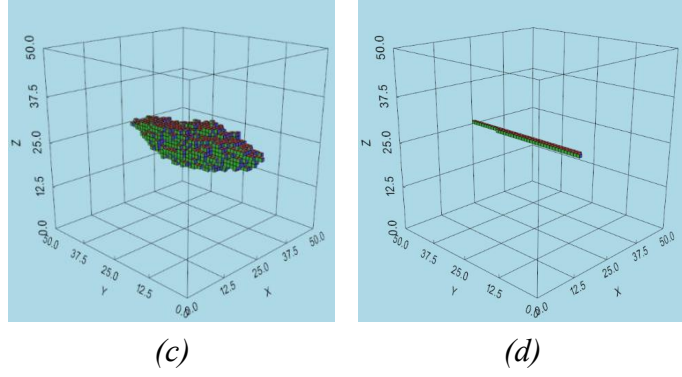

*(c)*          *(d)*

*Fig. 2. Nucleus shape developed under constant driving force* $dg_{bulk} = -\Delta\mu$

*(a)* $dg_{bulk} = 6.29909 \times 10^{-20}$ J $(E_3 < dg_{bulk} < E_{isol})$

*(b)* $dg_{bulk} = 5.74503 \times 10^{-20}$ J $(E_2 < dg_{bulk} < E_3)$

*(c)* $dg_{bulk} = 4.88454 \times 10^{-20}$ J $(E_1 < dg_{bulk} < E_2)$

*(d)* $dg_{bulk} = 2.59769 \times 10^{-20}$ J $(E_1 < dg_{bulk} < E_2)$

From Fig. 2 one may see that the homogeneous nucleation from single quasiatom becomes practically impossible if the driving force per atom is less than about $2.56 \times 10^{-20}$ J - this marginal energy belongs to the interval $(E_1 < dg_{bulk} < E_2)$. Gradual decrease of the bulk driving force leads to the shape change of crystallizing cluster from more-or-less equiaxial to needle-like. Note that this gradual shape transformation does not include quasi-planar (disc-like) shape. We believe, this is because of proximity between $E_2$ and $E_3$. In section IV and later, we will see that driving forces leading to needles (fibers) can be kept constant or nearly constant self-consistently, due to change of supersaturation during dissolution in limited volumes.

### III.2. Heterogeneous nucleation from the pre-existing rectangular clusters.

If we start simulation from some cluster, instead of single quasiatom, the minimal driving force, sufficient for further growth, becomes less, with increasing initial size. If the initial atomically flat facets are "spoiled" randomly ("shabby"), then the threshold driving force becomes even lower. Some examples:

#### III.2.a. *Pre-existing clusters with atomically flat facets.*

Growth of preexisting cluster becomes possible at $dg_{bulk} = 0.1 \times E_1$, if it's initial size is as big as $19 \times 19 \times 19$ periods. In Fig. 3a we demonstrate an example of size and shape evolution of the pre-existing cluster $10 \times 10 \times 10$ under driving force $dg_{bulk} = 3.8 \times 10^{-21}$ J after 1000000 attachment/detachment attempts (or upon the cluster reaching the sample boundaries).

#### III.2.b. *Pre-existing clusters with atomically flat facets.*

The deviations from flatness make easier both growth and shrinkage of the preexisting cluster. In Fig. 3b we demonstrate the evolution of initial $10 \times 10 \times 10$ cluster with roughened facets under the same driving force. Note that the initial cluster $9 \times 9 \times 9$ with roughened facets at the same driving force appears to be under critical, it starts shrinking and finally decomposes.

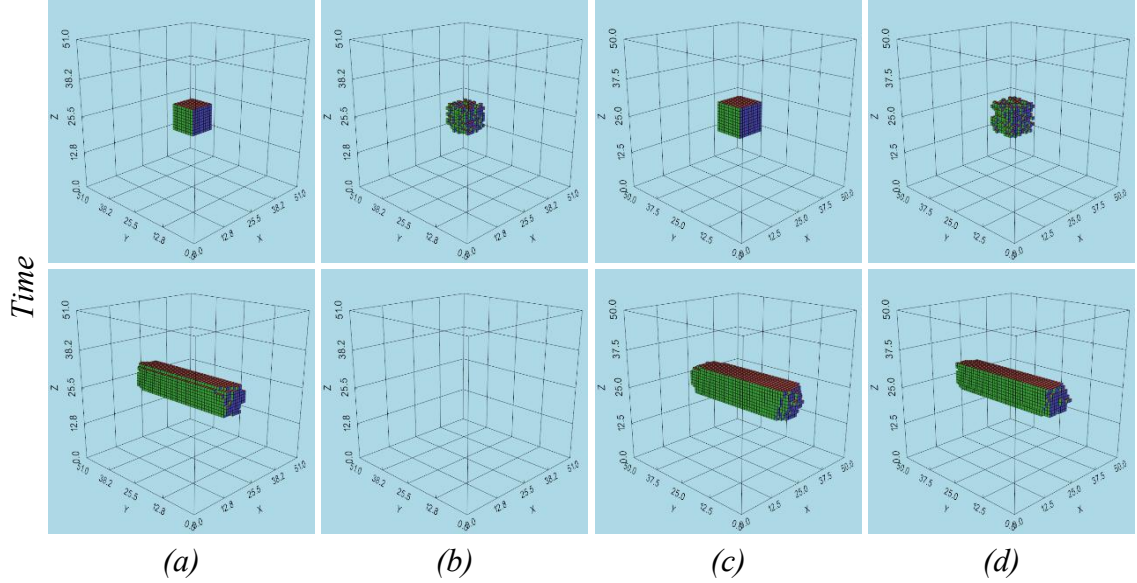

*Fig. 3. Evolution of clusters with different initial morphologies. Two scenarios are presented: (a, b) a $9 \times 9 \times 9$ cluster and (c, d) a $10 \times 10 \times 10$ cluster. For both cases, the driving force is $dg_{bulk} = 3.8 \times 10^{-21}$ J. Panels (a) and (c) demonstrate the evolution of clusters with atomically flat facets, while (b) and (d) show the evolution of clusters with initially "shabby" facets.*

## IV. Dissolution of initially rectangular crystal in the undersaturated medium at fixed bulk driving force

$$\Delta\mu = - kT \ln\left(\frac{C}{C_{eq}}\right) = const > 0 \; (dg = -\Delta\mu < 0)$$

Here we shortly analyze the case opposite to Section III. In this case the driving force of crystallization is negative, and the pre-exiting crystal should dissolve. We are interest in the shape evolution during such dissolution.

We start with the pre-exisiting parallelepiped $30 \times 30 \times 30$. At each step we make the same two attempts by Metropolis algorithm as in previous section, but now the driving force has opposite sign. We explored the following cases:

$$0 < \Delta\mu < E_1$$
$$E_1 < \Delta\mu < E_2$$
$$E_2 < \Delta\mu < E_3$$
$$E_3 < \Delta\mu < E_{isol}$$

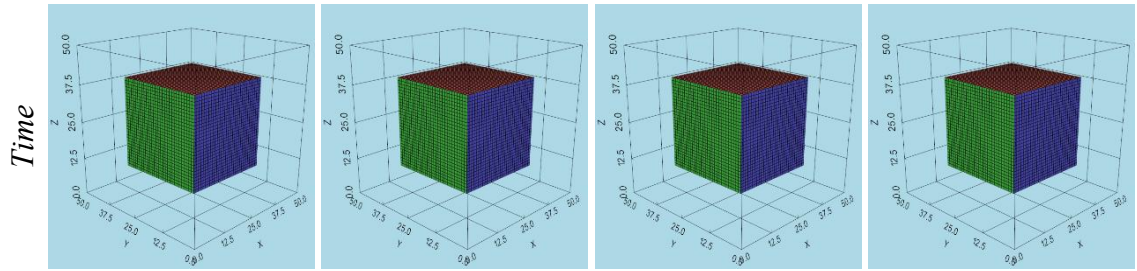

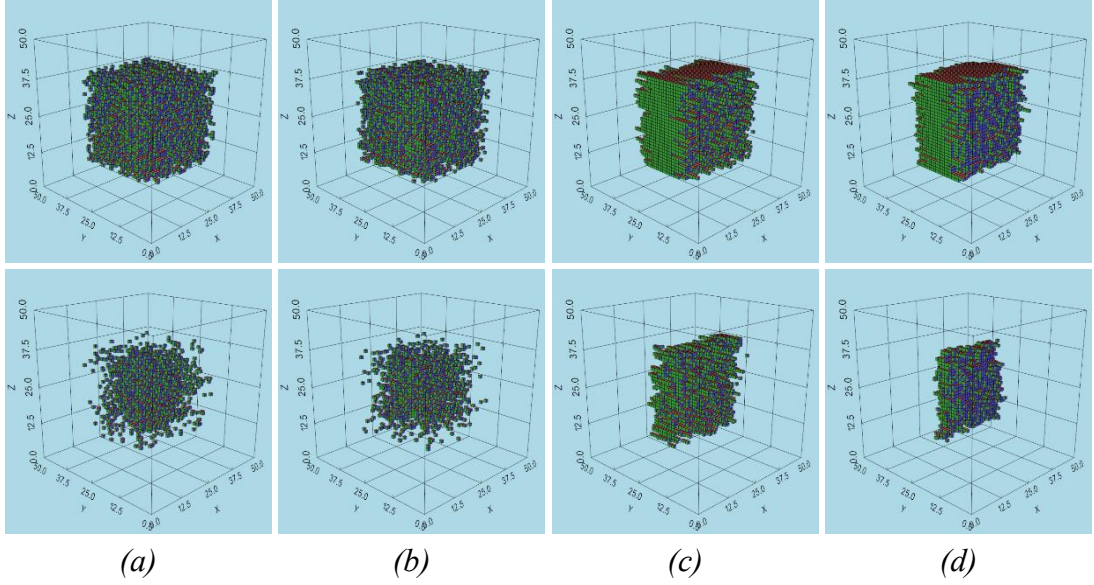

*(a)* *(b)* *(c)* *(d)*

*Fig. 4. Anisotropic dissolution kinetics with driving force belonging to 4 energetic intervals.*

*Case (a):* $E_3 < \Delta\mu < E_{\text{isol}}$, $dg_{\text{bulk}} = -6.30 \times 10^{-20}$ J – *shrinking proceeds fast and more or less isotropically.*

*Case (b):* $E_2 < \Delta\mu < E_3$, $dg_{\text{bulk}} = -5.745 \times 10^{-20}$ J.

*Case (c):* $E_1 < \Delta\mu < E_2$, $dg_{\text{bulk}} = -2.27 \times 10^{-20}$ J.

*Case (d):* $0.0 < \Delta\mu < E_1$, $dg_{\text{bulk}} = -9.48 \times 10^{-21}$ J.

One may see that the sample may become fibrous not only during crystallization (growth in supersaturated solution), but as well during melting (shrinking in undersaturated medium).

## V. Movement of solid liquid/interface at constant in time super- and under-saturations

We take, as an initial configuration, a crystalline parallelepiped occupying half of sites along axis X (or along axis Y, or along axis Z - in total, three initial configurations) and an along two other directions.

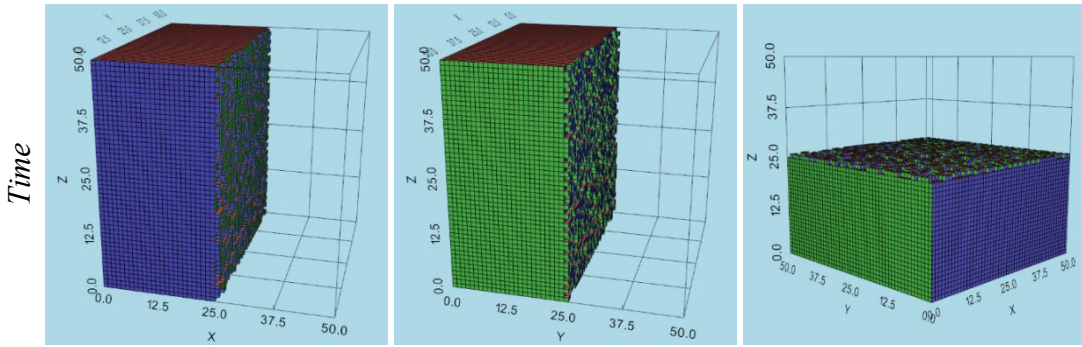

*Time*

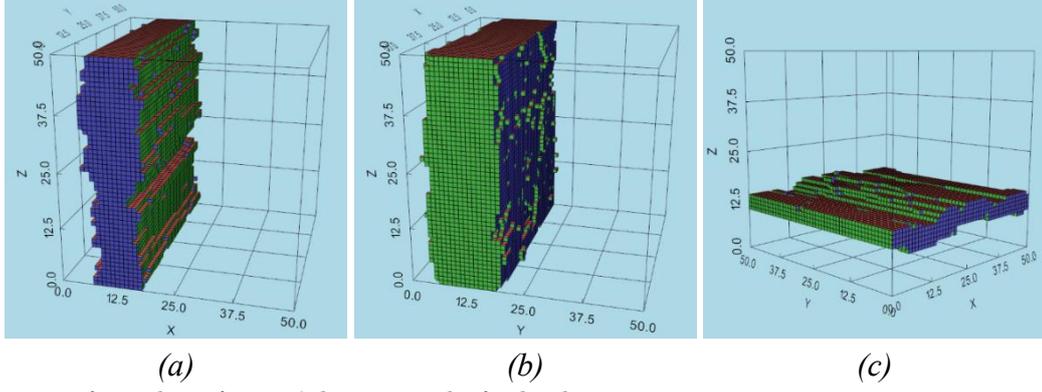

*(a)*                  *(b)*                  *(c)*

*Fig. 5. Movement of crystal interfaces in 3 directions under fixed undersaturations_*

*(a) Direction ⟨100⟩: $dg_{bulk} = -2.63457 \times 10^{-20}$ J*

*(b) Direction ⟨010⟩: $dg_{bulk} = -1.89688 \times 10^{-21}$ J*

*(c) Direction ⟨001⟩: $dg_{bulk} = -3.00342 \times 10^{-20}$ J*

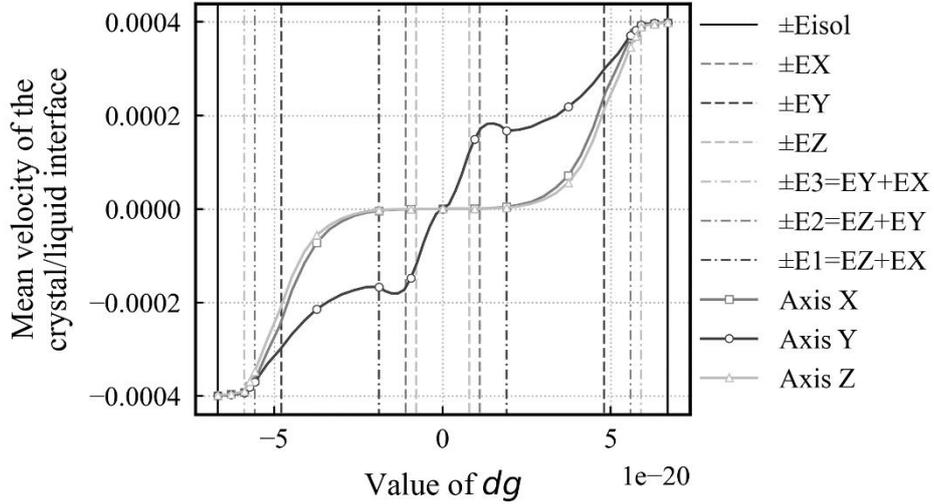

*Fig. 6. Dependences of crystal interface velocity V on the driving force dg along axes X,Y,Z: Plot VX(dg = −Δμ), Plot VY(dg = −Δμ), Plot VZ(dg = −Δμ).*

Let us compare the dependences at Fig. 6 with the characteristic energies $E_1$, $E_2$, $E_3$, $E_X$, $E_Y$, $E_Z$. As expected, the dependence of velocity on driving force is far from linear, if we explore the broad interval of the driving force. First of all, as expected, the fastest growth at fixed driving force is observed in the direction along axis $Y$, perpendicular to facets ⟨010⟩. This is because the increase of energy due to attachment to ⟨010⟩ terrace is $2 \times E_1$ and it is significantly less than the energies for other directions ($2 \times E_2$ and $2 \times E_3$).

Unexpected for us was the small energetic interval between $E_X$ and $E_1 = E_Z + E_X$. We believe that if the driving force above $E_Z$ and $E_X$, the terraces ⟨010⟩ may grow laterally very fast, and form the atomically flat ⟨010⟩ surface. Since the dg is still less than E1, we should wait for some time to reach the large enough fluctuation, necessary to attach viable 2D-nucleus of the next ⟨010⟩ atomic layer.

## VI. Evolution of pre-existing cluster in undersaturated medium with fixed volume and self-regulated driving force (concentration)

In previous sections, we treated the bulk driving force per quasiatom, $dg = -\Delta\mu$, as a fixed external value, constant in time during growth or shrinkage. Here, we will switch to a model with self-regulated concentration C of quasiatoms in the surrounding medium, leading to a corresponding self-regulated driving force, $dg = -\Delta\mu = kT\ln\left(\frac{C}{C_{eq}}\right)$.

Namely, we fix the total number of sites for quasiatoms in our model system, $N_{total}$, treating this total system as a lattice gas, part of which forms single (so far) crystalline cluster. This number cannot change during the process of growth or shrinkage. We also fix the initial number of sites belonging to the crystalline cluster, $N_{cryst0}$, and the initial concentration, $C_0$, of quasiatoms in the surrounding medium. Thus, the initial number of quasiatoms in the medium is $N_{gas0} = C_0 \cdot (N_{total} - N_{cryst0})$.

The total number of quasiatoms in both phases remains constant: $N_{qa} = N_{gas} + N_{cryst} = N_{gas0} + N_{cryst0} = C_0 \cdot (N_{total} - N_{cryst0}) + N_{cryst0} = $ const.

During the process, after each successful attachment or detachment of a quasiatom to (or from) the crystalline cluster, the concentration of quasiatoms is recalculated as $C = N_{gas}/(N_{total} - N_{cryst}) = (N_{qa} - N_{cryst})/(N_{total} - N_{cryst})$. This recalculated concentration yields a new driving force: $dg = -\Delta\mu = kT\ln\left(\frac{C}{C_{eq}}\right)$. The rest of the algorithm remains the same.

Typical results of modeling are shown below

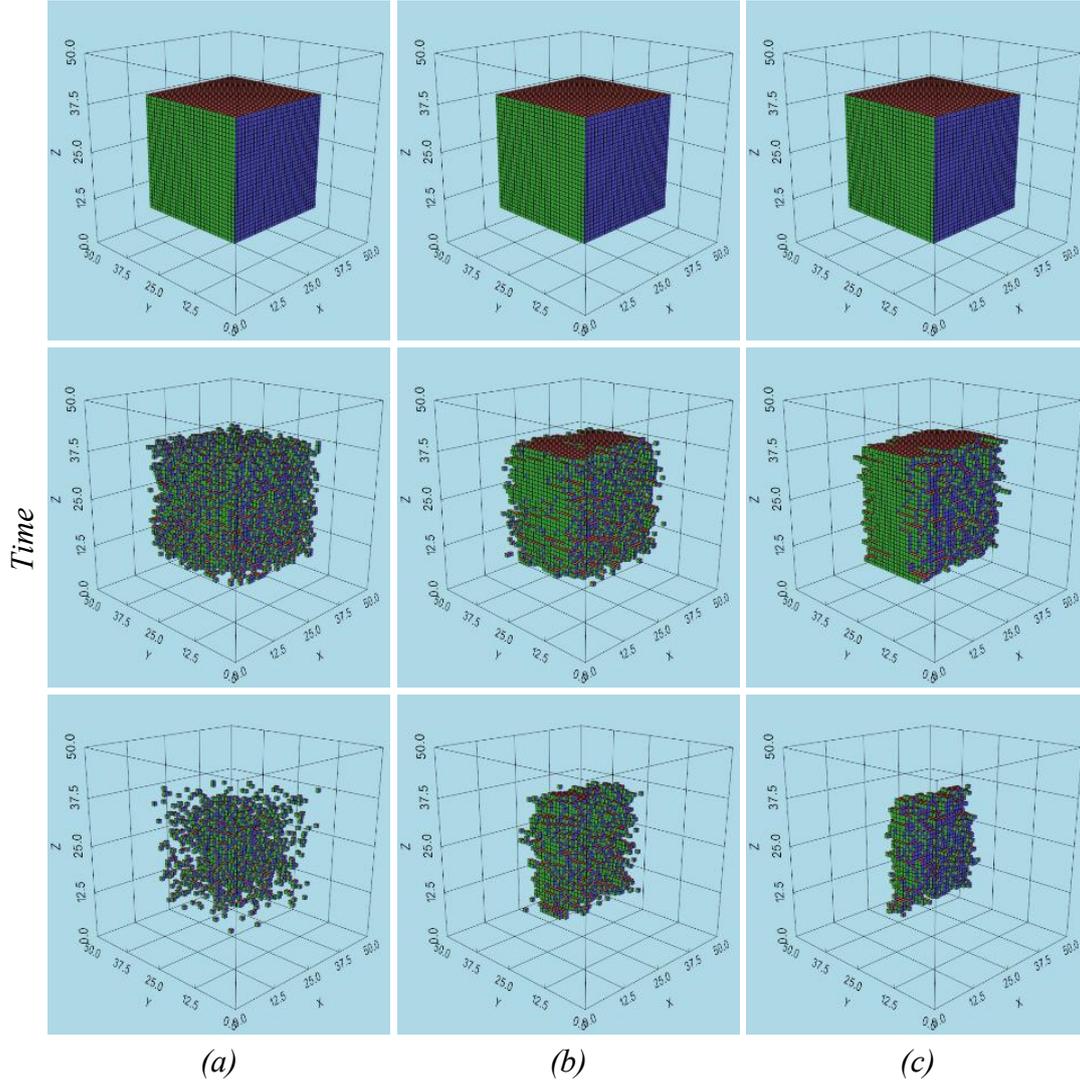

*Time*

(a)            (b)            (c)

*Fig. 7. Evolution of nanoparticles in a limited volume with self-regulated under/supersaturation without stirring – leading to very rough surfaces. In all cases $N_{total} = 1e22$*

*Case (a):* $C_0 = 6.17247 \times 10^{-14}$, $(dg_{bulk} =- 5.90471 \times 10^{-20}$ J$)$. *"Time" till full dissolution 27750 TRI.*

*Case (b):* $C_0 = 1.14565 \times 10^{-11}$, $(dg_{bulk} =- 3.74111 \times 10^{-20}$ J$)$. *"Time" till full dissolution 62543 TRI.*

*Case (c):* $C_0 = 9.71066 \times 10^{-9}$, $(dg_{bulk} =- 9.48439 \times 10^{-21}$ J$)$. *"Time" till full dissolution 99949 TRI.*

From Fig. 7 one may see, that the initial undersaturation (depletion) of the surrounding medium in limited volume may lead to formation of fibrous morphologies even without any stirring.

So far, in this section, we treated the initial undersaturation and total number of site $N_{total}$ so large that eventually the initial crystalline cluster was dissolved. Now we choose the much less volume of surrounding medium, so that the crystalline cluster survives, and after partial dissolution system comes to equilibrium. Note that the size of cluster (number of sites $N_{cryst}$) and concentration of quasi-atoms in gas state tends to asymptotic (expected as an equilibrium value) much faster than the shape and the corresponding surface energy of the cluster (Fig. 8a and Fig. 8b). The geometry of cluster in equilibrium state (Fig. 8c) is tending also to some asymptotics which is expected to coincide with the Wulff's rule.

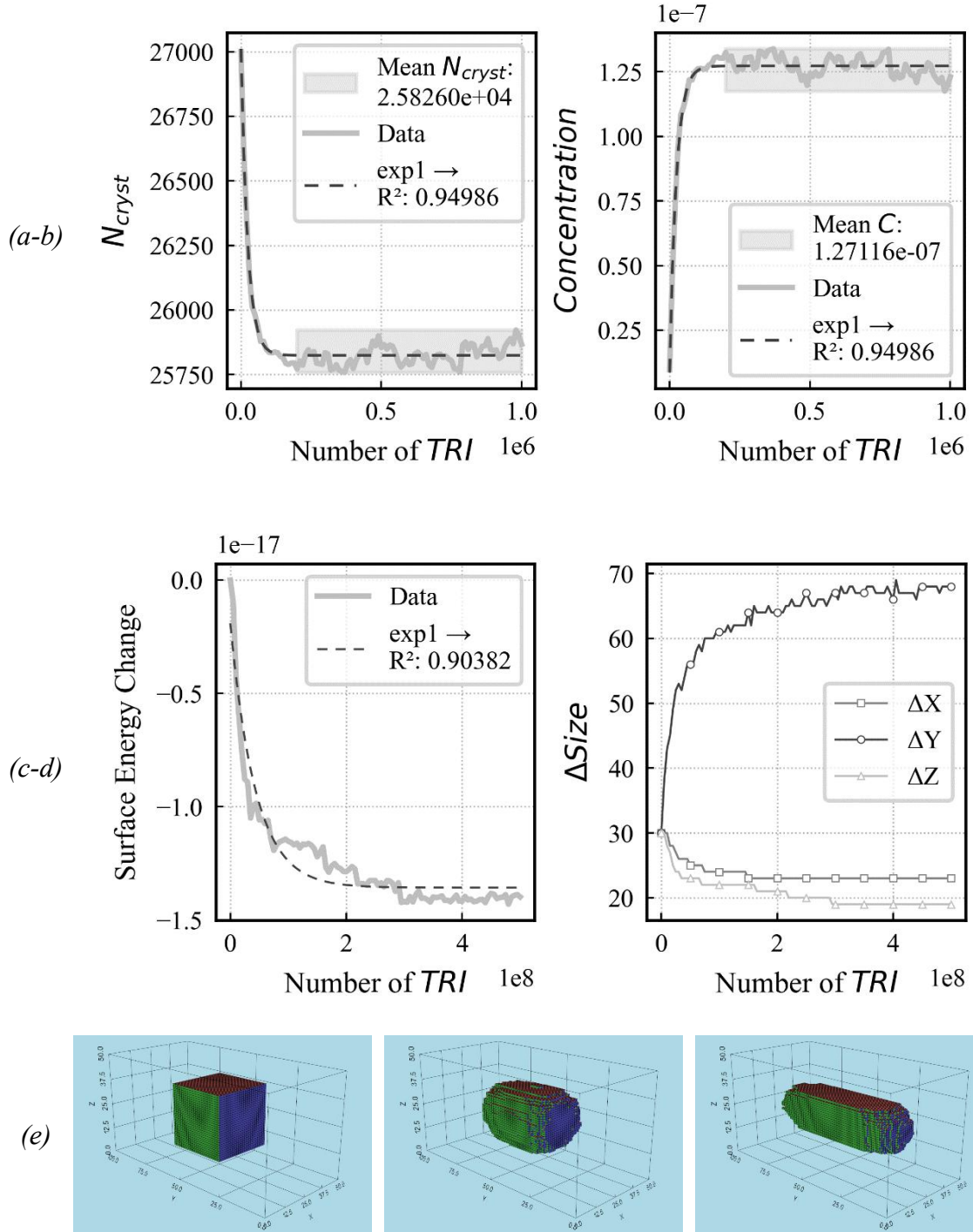

Fig. 8. Self-regulation of the crystalline cluster's size and shape.

(a) Time dependence of $N_{\text{cryst}}$ at $N_{\text{cryst0}} = 27000$, $C_0 = 9.71066 \times 10^{-9}$, $N_{\text{total}} = 10^{10}$, $t_{\max} = 10^6\ TRI$.

(b) Time dependence of concentration. The asymptotic self-regulated concentration is higher than the equilibrium value. As shown in Appendix A, this deviation corresponds to the Gibbs-Thomson effect for nanoparticles in equilibrium with the surrounding medium. Specifically, the size correction factor, $\exp\left(4\gamma^{ef}\Omega^{2/3}/kTN_{\text{cryst}}^{1/3}\right)$, for our data gives 1.296, and the asymptotic concentration is $1.242 \times 10^{-7}$.

(c) Time dependence of the crystalline surface energy change (with respect to the initial value) at $N_{\text{cryst0}} = 27000$, $C_0 = 9.71066 \times 10^{-9}$, $N_{\text{total}} = 10^{10}$, $t_{\max} = 5 \times 10^8\ TRI$.

(d) Evolution of sizes $\Delta X, \Delta Y$, and $\Delta Z$ (here, for example, $\Delta X = X_{\max} - X_{\min}$). Note that the scales for $X, Y$, and $Z$ are different. The asymptotic ratio of sizes deviates significantly from the Wulff rule. For an ideal parallelepiped, one would expect:

$$(\Delta Y \cdot a_{1y})/(\Delta X \cdot a_{1x}) = \gamma(010)/\gamma(100) = 1.32$$

$$(\Delta Z \cdot a_{1z})/(\Delta X \cdot a_{1x}) = \gamma(001)/\gamma(100) = 0.54$$

$$(\Delta Y \cdot a_{1y})/(\Delta Z \cdot a_{1z}) = \gamma(010)/\gamma(001) = 2.45$$

Instead, we obtained 0.87, 0.66, and 1.33, respectively. Notably, the minimal deviation from the Wulff rule is observed for directions perpendicular to the elongation direction. We attribute these deviations to the non-rectangular shape of the particles. In any case, the observed shape corresponds to the minimal surface energy.

(e) Change in shape during self-regulation of size and composition.

As can be seen from Fig.8b, the asymptotic self-regulated concentration is higher than the equilibrium value. As shown in Appendix A, this deviation corresponds to the Gibbs-Thomson effect for nanoparticles in equilibrium with the surrounding medium. Specifically, the size correction factor, $exp\left(4\gamma^{ef}\Omega^{2/3}/kTN_{cryst}^{1/3}\right)$, for our data gives 1.296, and the asymptotic concentration is $1.24 \times 10^{-7}$.

## VII. Introduction of isotropic ballistic terms into detachment probability

To take intensive stirring into account, we add one more (third) opportunity to the two existing opportunities of the thermoactivated elementary events - attachment and detachment of quasiatom with Boltzmann probability. This third opportunity is a possibility of ballistic (not thermo-activated) detachment of quasiatom, which we may vary but (in this Section) we take as the same for all directions.

We expect that the ballistic events will detach additional quasiatoms from the crystalline cluster, leading to supersaturation, with (if not too small and if not too large) may lead to preferential growth along Y direction. This should lead to conversion of cube into elongated parallelepiped (belt or fiber). And indeed, we observe very similar behavior at Fig. 9.

Namely, the particle is elongating along axis $Y$ and finally tending to asymptotic elongation, showing the influence of ballistic factor on the shape evolution.

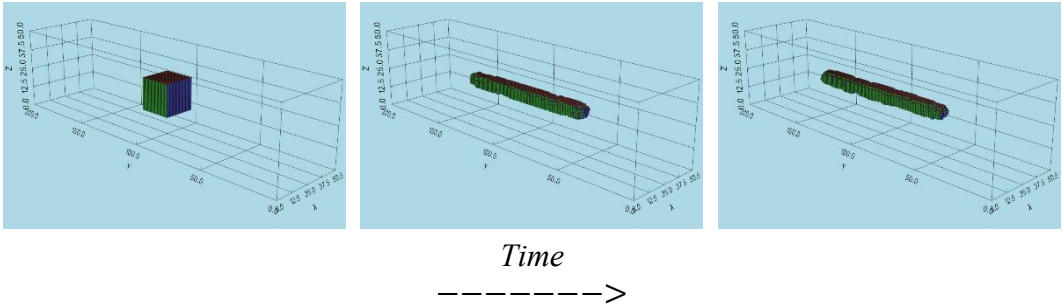

*Time*
-------->

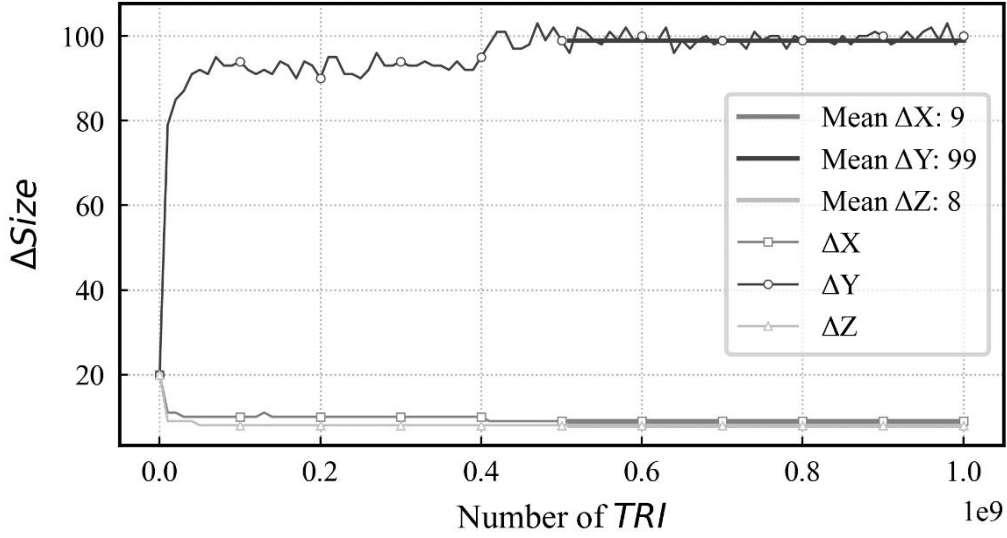

*Fig. 9. Transformation of cube into nano-belt under ballistic events (corresponding to intensive stirring).*

*Parameters:* $N_{\text{cryst0}} = 8000$, $C_0 = 9.58767 \times 10^{-9}$, $N_{\text{total}} = 10^{10}$, $P_{\text{ballistic}} = 0.05$, $t_{\text{max}} = 10^9$ *TRI*.

## VIII. Introduction of anisotropic ballistic terms into detachment probability

Here we introduce the dependence of ballistic probability on the energy of detaching quasiatom, which makes the ballistic events anisotropic. Namely, we took

$$P_{\text{ballistic}} = P_0 \cdot \left(1.0 - {dE_{\text{surface}}}/{E_{\text{isol}}}\right)$$

Here $dE_{\text{surface}}$ is a change of surface energy due to ballistic detachment of quasiatom. If quasiatom would be taken from the bulk, the new-formed vacancy in the bulk would increase the surface energy by the energies of 6 facets equal to $E_{\text{isol}}$. According to our equation, probability of such event is taken to be zero. The result of simulation appears to be qualitatively close to previous section, except the magnitude of elongation.

If one changes the magnitude of parameter $P_0$ from $P_0 = 0.17$ to $P_0 = 0.085$ (twice less), the asymptotical length decreases by approximately 25 %

On the other hand, using alternative equation for $P_{\text{ballistic}}$ with the square of the ratio $\left({dE_{\text{surface}}}/{E_{\text{isol}}}\right)$ instead of linear dependence, we get the asymptotic $\Delta Y$ approximately 38 instead of 67.

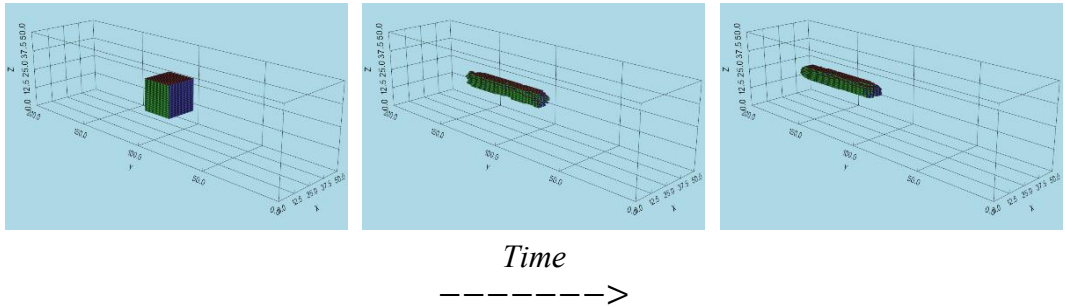

*Time*
-------→

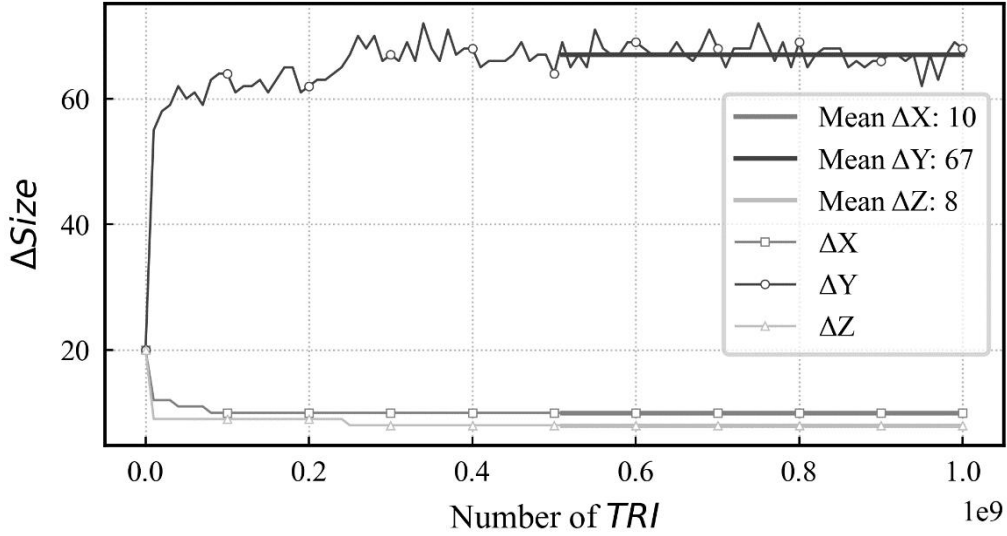

*Fig. 10. Transformation of cube into nano-belt under ballistic events (corresponding to intensive stirring).*

*Parameters:* $N_{cryst0} = 8000$, $C_0 = 9.58767 \times 10^{-9}$, $N_{total} = 10^{10}$, $P_0 = 0.17$, $t_{max} = 10^9$ *TRI*.

## IX. Account of many particles interacting via mean field under stirring

To see the possible effect of ripening, we now switch to simulation of the ensemble of particles which share the common volume and may realize an exchange of quasiatoms via the mean-field. Therefore, in our programming we introduced the array of N0 initial particles with randomly modified surfaces and simulated the possible thermal attachments and detachments, as well as ballistic detachments for all of them. We chose N0=100 and the parameter Pb for ballistic probability as 0.1. As shown at Fig. 11, we observed (a) the decrease of number of remaining particles (11a), (b) increase of concentration with subsequent tending to almost constant value (11b), (c) growth of total surface energy (instead of tending to minimum for closed systems) (11c), (d) constant elongation of mean Y-length in (010) direction, and tending to minimal asymptotic values in X- and Z-directions (11d), (e) growth of size Y dispersion, (f) linear time growth of mean squared (010)-size

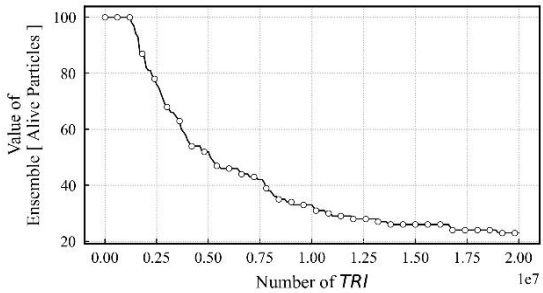

*(a)*

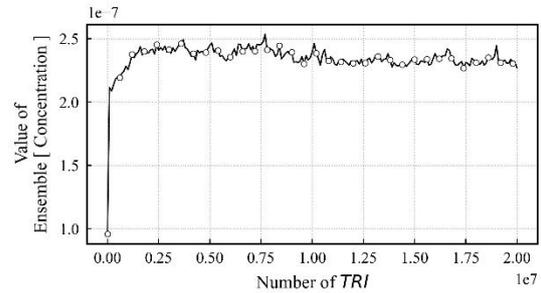

*(b)*

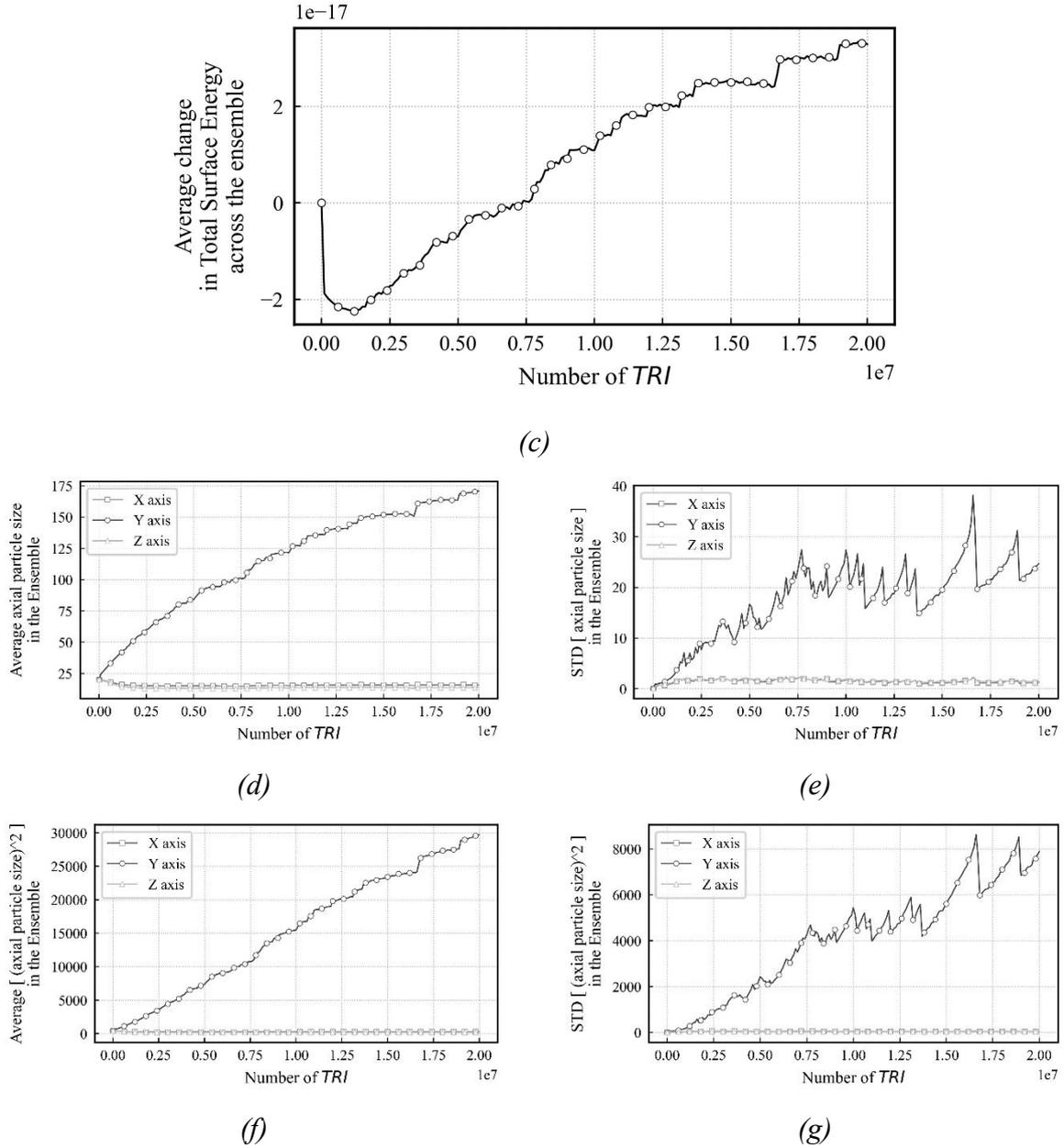

*(c)*

*(d)*         *(e)*

*(f)*         *(g)*

*Fig. 11. Time dependences (in trial iteration units – TRI) of (a) number of remaining particles, (b) mean-field concentration, (c) total surface energy, (d) mean values of sizes, (e) standard deviation of size, (f) mean squared sizes, (g) standard deviations of squared sizes.*

Parameters: $P_b = 0.05$, $N_0 = 100$, $T = 300\,K$.

## X. Conclusions

(1) Transformation of more or less equiaxial particles of $V_2O_5$ into fibers or belts can be well described within the Monte Carlo simulation scheme (with account of terrace-ledge-kink structure of all surfaces) by introducing the ballistic (athermal) elementary events corresponding to intensive stirring.

(2) At first, we study the behavior of single particle with anisotropic surface energies under fixed thermodynamic driving force (change of chemical potential due to crystallization) which is kept constant.

(3) Not any high supersaturation (high driving force) may lead to large anisotropy and fibrous structures. Emergency of very elongated structures happens if the driving force belongs to some interval – if it is big but not too big, otherwise the structure becomes more-or-less equiaxial)

(4) Elongation of single particle becomes realistic in limited volume providing the self-regulation of composition of the surrounding medium in the case of ballistic detachment.

(5) Without detachment the self-regulation leads to equilibrium asymptotic composition, to minimal energy and to almost equilibrium shape (with account of Gibbs-Thomson capillary effect for almost rectangular particle)

(6) Namely, intensive ballistic detachments of atoms lead to high supersaturations leading to non-linear highly anisotropic growth – thus, cubes transforms into belts.

(7) Very slow formation of fibrous structures without stirring, maybe, can be described as anisotropic shrinkage.

(8) Late stages of evolution for the ensemble case show some features characteristic for ripening:

(a) Number of particles decreases with time.
(b) Concentration tends to some asymptotic value with time
(c) Mean length along (010)-direction increases according to power law, but ½ instead of 1/3.

On the other hand, some features of "antiripening" are also observed:

(d) Total surface energy increases instead of tending to minimum.
(e) Mean sizes along (100) and (001) directions tend to constants instead of growth.

**Acknowledgments**

Authors acknowledge the grant of Ministry of Education and Science of Ukraine 0125U001491, corresponding author acknowledges the hospitality of MSE Faculty of Technion as a visiting professor during February-July 2025.

**Appendix A. Optimization of cluster shape and composition around cluster.**

We will assume cluster to keep rectangular form, so that shape is parametrized by two shape factors $\varphi_y = b/a$, $\varphi_z = c/a$. Solution around cluster is described as a lattice gas with number of sites Ngas=Ntotal -Ncrystal, concentration C=Nquasiatoms/Ngas. We should minimize the free energy

$$F = 2\gamma(100)bc + 2\gamma(010)ca + 2\gamma(001)ab + Eisol * C * Ngas - kT * \ln\left(\frac{Ngas!}{(C*Ngas)!((1-C)Ngas)!}\right) \quad (A1)$$

Constraints of matter conservation:
Ngas=Ntotal-Ncrystal => dNgas=-dNcryst,  (A2)
CNgas+Ncryst=C0*Ngas0+Ncryst0=const =>
$(Ntot - Ncryst)dC + (1 - C)dNcryst = 0$ =>
$$\frac{dNcryst}{dC} = -\frac{Ntot - Ncryst}{1 - C}. \quad (A3)$$

Sizes of cluster are expressed in terms of shape factors:
$$a = (Ncryst\Omega)^{1/3} * \varphi_y^{-1/3}\varphi_z^{-1/3},$$
$$b = (Ncryst\Omega)^{1/3} * \varphi_y^{2/3}\varphi_z^{-1/3},$$
$$c = (Ncryst\Omega)^{1/3} * \varphi_y^{-1/3}\varphi_z^{2/3}$$

Thus, after using Stirling approximation,
$$F = (Ncryst(C) * \Omega)^{2/3} * (\gamma(100)\varphi_y^{1/3}\varphi_z^{1/3} + \gamma(010)\varphi_y^{-2/3}\varphi_z^{1/3} + \gamma(001)\varphi_y^{1/3}\varphi_z^{-2/3}) + CNgas(C)Eisol + kT * (Ntotal - Ncrystal(C))(ClnC + (1 - C)\ln(1 - C)) = F(C,\varphi_y, \varphi_z) \quad (A4)$$

Then the set of 3 equations $\frac{\partial F}{\partial \varphi_y} = 0$, $\frac{\partial F}{\partial \varphi_z} = 0$, $\frac{\partial F}{\partial C} = 0$ gives:

$\varphi_y^{opt} = \gamma(010)/\gamma(100)$, $\varphi_z^{opt} = \gamma(001)/\gamma(100)$,  (A5)

$C^{opt} = C^{equil} = \exp\left(-\frac{Eisol}{kT}\right) * \exp\left(\frac{4\gamma^{ef}\Omega^{2/3}}{kT*N_{cryst}^{1/3}}\right)$  (A6)

with $\gamma^{ef} = (\gamma(100) * \gamma(100) * \gamma(100))^{1/3}$  (A7)

Equation (A6) is actually a generalization of Gibbs-Thomson relation on the anisotropic case. Strictly speaking, in case of self-regulated cluster evolution in limited volume, equatio(A6) may be treated as a transendent equation for C, since Ncryst depends on C (see (A3))